\documentclass[preprint,showpacs,preprintnumbers,amsmath,amssymb,showkeys]{revtex4}
\usepackage{txfonts}
\usepackage{amsmath}
\usepackage{graphicx}
\usepackage{dcolumn}
\usepackage{bm}
\usepackage{overpic}

\begin{document}
\title{Maximizing Entropy Yields Spatial Scaling in Social Networks}
\author{Yanqing Hu$^1$\footnote{yanqing.hu.sc@gmail.com}, Yougui Wang$^1$, Daqing Li$^2$, Shlomo Havlin$^2$, Zengru Di$^1$\footnote{zdi@bnu.edu.cn}}
 \affiliation{1. Department of Systems Science, School of Management and Center for Complexity
 Research, Beijing Normal University, Beijing 100875, China
\\2. Department of Physics, Bar-Ilan University, Ramat-Gan 52900, Israel
 }

\date{\today}

\begin{abstract}
In addition to the well known common properties such as small world
and community structures, recent empirical investigations suggest a
universal scaling law for the spatial structure of social networks.
It is found that the probability density distribution of an
individual to have a friend at distance $r$ scales as $P(r)\propto
r^{-1}$. The basic principle that yields this spatial scaling
property is not yet understood. Here we propose a fundamental origin
for this law based on the concept of entropy. We show that this
spatial scaling law can result from maximization of information
entropy, which means individuals seek to maximize the diversity of
their friendships. Such spatial distribution can benefit individuals
significantly in optimally collecting information in a social
network.

\end{abstract}



\maketitle

Social networks structure is found to be important since it leads to
deep insights about how people interact and how social relations
evolve \cite{Lai Ying-Cheng, BA model, first issue b, Newman-review,
Givens-Newman, Song fractal 1, JShao, Kleinberg-covergence,
wattsSCI,Cellphone,Alx1,Alx2,humaninteraction,PhoneInfections}. It
has been found that social networks possess common properties such
as small-world \cite{first issue b} and community structure
\cite{Givens-Newman}. Recently, geographical properties of social
networks have attracted much attention \cite{first issue b,the
oldest experiment, new experiment, navigation brief nature,
navigation full, power-law networks Kleinberg search, power-law
networks search, News, Use Kleinberg search, Analyzing kleinberg,
Kleinberg hierarchical model, Oskar licentiate thesis, Renaud
Lambiotte,Goldenberg}. Several empirical studies have analyzed the
distribution of distances between friends in real social networks.
Liben-Nowell \textit{et al.} explored the geographic properties in
decentralized search within a large, online social network \cite{Use
Kleinberg search}. They used data from the LiveJournal online
community with about 500,000 members, in which their state and city
of residence, as well as a list of their LiveJournal friends are
available. They found that the probability density function (PDF),
$P(r)$, of an individual having a friend at a geographic distance
$r$ is about $P(r)\propto r^{-1}$ (see supplementary I). Almost at
the same time, Adamic and Adar have also found the same phenomenon
\cite{power-law networks Kleinberg search}. They investigated a
relatively small social network, the Hewlett-Packard Labs email
network. In this work, the PDF of the distance is also found to
scale as $P(r)\propto r^{-1}$. More recently, Lambiotte \textit{et
al.} investigated a large mobile phone communication network
\cite{Renaud Lambiotte}. The network consists of 2.5 million mobile
phone customers that have placed 810 million communications, for
whom they have the geographical home location information. Their
empirical results show that the mobile phone communication network
has the same scaling properties in the spatial structure. They found
that probability of two nodes $(u$ and $v)$ to have a long range
connection of length $r(u,v)$ is $Pr(u,v)\propto r(u,v)^{-2}$. For
$2$-dimensional space, the number of nodes which have distance $r$
from a given node is proportional to $r$. This implies that the PDF
of an individual to have a friend at distance $r$ is $P(r)\propto
r\cdot r^{-2}=r^{-1}$. Very recently, Goldenberg and Levy
investigated several large online communities, and also detected the
same spatial scaling phenomenon \cite{Goldenberg}. From the above
empirical investigations, one can conclude that the PDF of having a
friend at distance $r$ is
\begin{equation}
P(r)\propto r^{-1}.\label{distribution}\end{equation}

Why does the spatial structure of our social networks possess this
kind of scaling property and how does it benefit us? Kleinberg has
proved that in a $d$-dimensional space, when the probability of
having a long range connection of length $r$ between $u$ and $v$ is
$Pr(u,v)\propto r(u,v)^{-d}$, the network is optimally navigated
\cite{navigation brief nature, navigation full,add1,add2}. For
$d$-dimensional lattice, the number of nodes that have the same
distance $r$ to a given node is proportional to $r^{d-1}$. So when
the network structure is optimal for navigability, the PDF of the
distance from a given node is $P(r)\propto r^{d-1}\cdot r^{-d}=
r^{-1}$ for all $d$. This spatial scaling property enables people to
send messages efficiently in minimal number of hops to \emph{all}
nodes of the system. However, social networks are usually not
constructed for the purpose of sending messages between
\emph{unrelated} individuals. Thus, there should be a fundamental
origin that governs the formation of the spatial scaling law, Eq.
(1).

Here we suggest that the origin of this scaling, Eq. (1), comes from
a general perspective based on the concept of entropy. We
hypothesize that human social behavior is based on gathering maximum
information through different activities. Making friends can be
regarded as a way of collecting information. To get optimal
information could be a general purpose for an individual that shapes
the social network architecture. We will show that a social network
based on Eq. (\ref{distribution}) is an optimal network which can
benefit people in collecting maximal information.

\section{Model}
To model a social system we use a toroidal lattice to denote the
world in which each node represents an individual. We assume that
each individual has a finite energy $w$ which can be represented by
the sum of distances between an individual and all his or her
friends,
\begin{equation}\sum_{v=1}^{m}r(u,v)=w,\label{energy-limition}\end{equation}
where $m$ is the number of direct links of node $u$. Eq.
(\ref{energy-limition}) implies that every node $u$ selects its long
range acquaintances $v$, one by one, until the total distance
reaches $w$.

\begin{figure}
\center
\includegraphics[width=7cm]{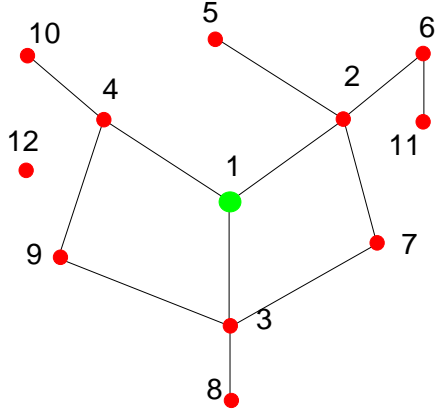}
\caption{The friends of node 1. Node 2, 3 and 4 are the friends of
node 1 which Eq. (2) yields that $d(1,2)+d(1,3)+d(1,4)=w$. The size
of the network is $n=12$ and the information sequence is
$\{2,3,4,5,6,7,7,8,9,9,10\}$ and the frequencies of all nodes are
$q_2=q_3=q_4=q_5=q_6=q_8=q_{10}=\frac{1}{11}$,
$q_7=q_9=\frac{2}{11}$, ${q_1}=q_{11}=q_{12}=0$. If one site is
reached several times when constructing the long range connections
from node 1 or from its nearest neighbors, the node will appear in
the node sequence and in Eq. (2) the same number of
times.}\label{example}
\end{figure}

The information that node $v$ brings to $u$ can be evaluated by
considering the information of node $v$ and all its neighbors. Thus,
the information that $u$ collects can be expressed by the sequence
of nodes as illustrated in Fig. \ref{example} and the entropy of the
whole sequence measures the amount of information. We assume that
all nodes are equivalent, so the information obtained by one node
can represent the information obtained by each of the other nodes.
Thus, our model for constructing a social network is
\begin{equation}\textit{Max}\,\,\varepsilon=-\sum^{n}_{i=1}q_i\log{q_i},\label{objective_function}\end{equation}
subjected to Eq. (\ref{energy-limition}). In Eq.
(\ref{objective_function}), $q_i$ denotes the frequency of node $i$
in the information sequence (see Fig. 1) and $n$ is the size of the
network. When $i$ is not a neighbor and not a next nearest neighbor
of $u$, $q_i=0$, and we define $q_i\log{q_i}=0$. Here, Eq.
(\ref{objective_function}) implies that the information entropy
$\varepsilon$ is determined by the sequence of friends and friends
of friends (For considering also friends of next nearest friends,
see supplementary IIA).

\section{Results}
Our optimization model (OM) is based on Eqs.
(\ref{energy-limition}) and (\ref{objective_function}) which
represent two competing processes. To maximize entropy (Eq.
(\ref{objective_function})), it is preferred to have friends at long
distances in order to explore new parts of the network and to obtain
more information. However the farther one goes he can have less
friends due to the finite energy limited by Eq.
(\ref{energy-limition}). Assuming the PDF of having a friend at
distance $r$ obeys
\begin{equation}P(r)\propto r^{-\alpha},\label{Power-law}\end{equation}
we can explore the value of $\alpha$ that yields maximum entropy
under the condition of Eq. (\ref{energy-limition}).

The optimization model is simulated on a toroidal lattice whose size
is $L\times L$ ($L=10000$ means that individuals can make friends in
a population of $10^8$) and lattice (`Manhattan') distance is
employed. Because toroidal lattice is a regular network and each
node has a unique index, we can calculate the lattice distance
between any pair of nodes and we do not need to construct the whole
network, enabling us to simulate very large lattices.

\begin{figure}
\center
\includegraphics[width=4.4cm,height=4cm]{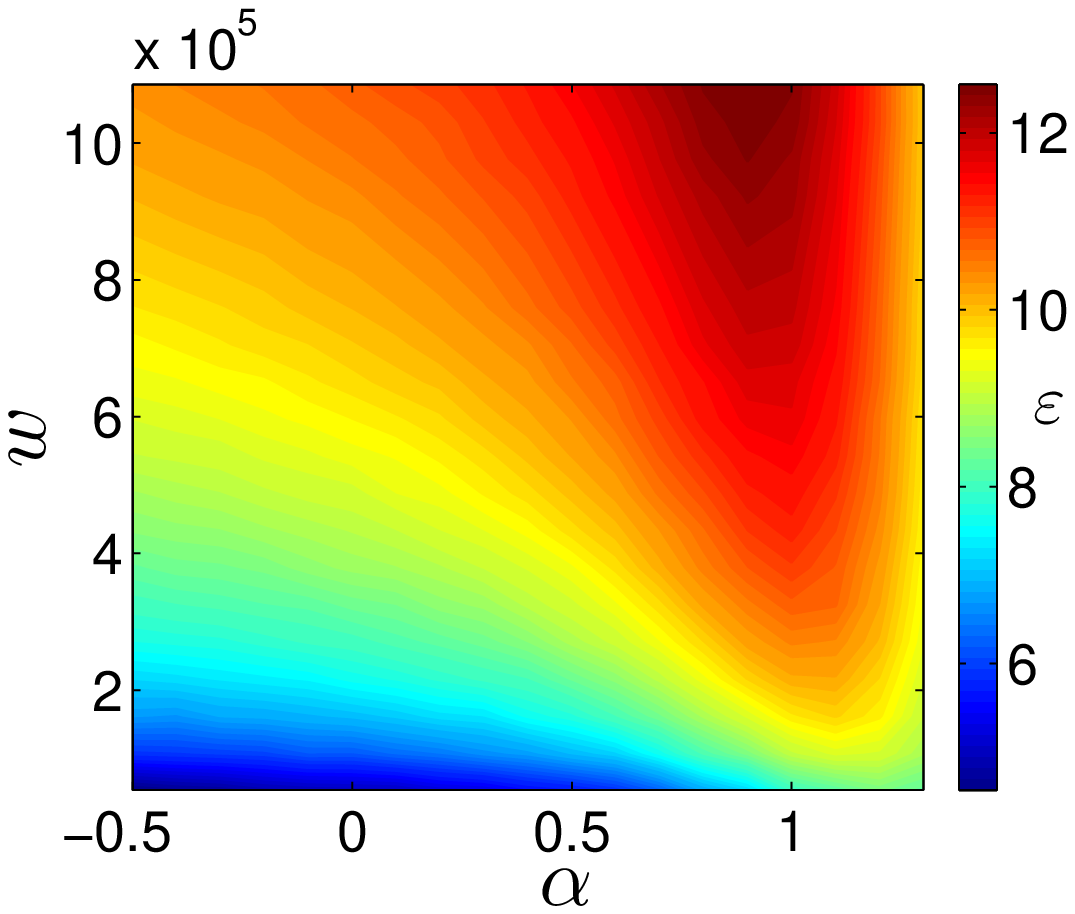}\includegraphics[width=4.4cm,height=3.8cm]{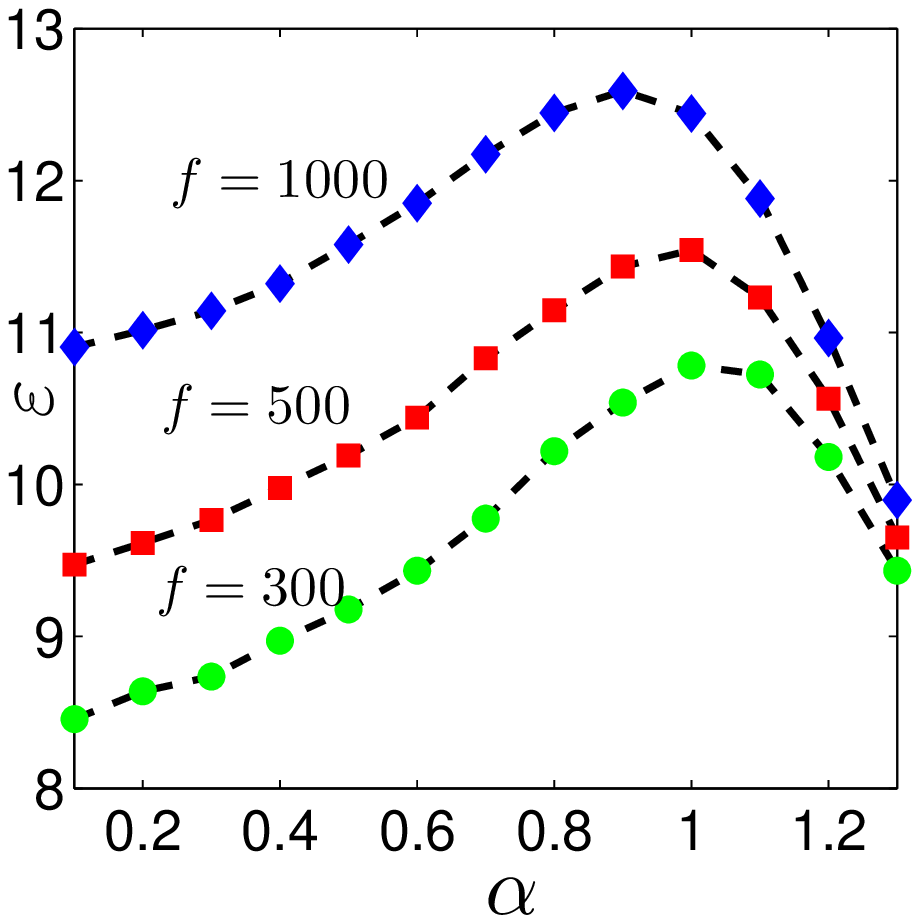}
\includegraphics[width=4.4cm,height=3.8cm]{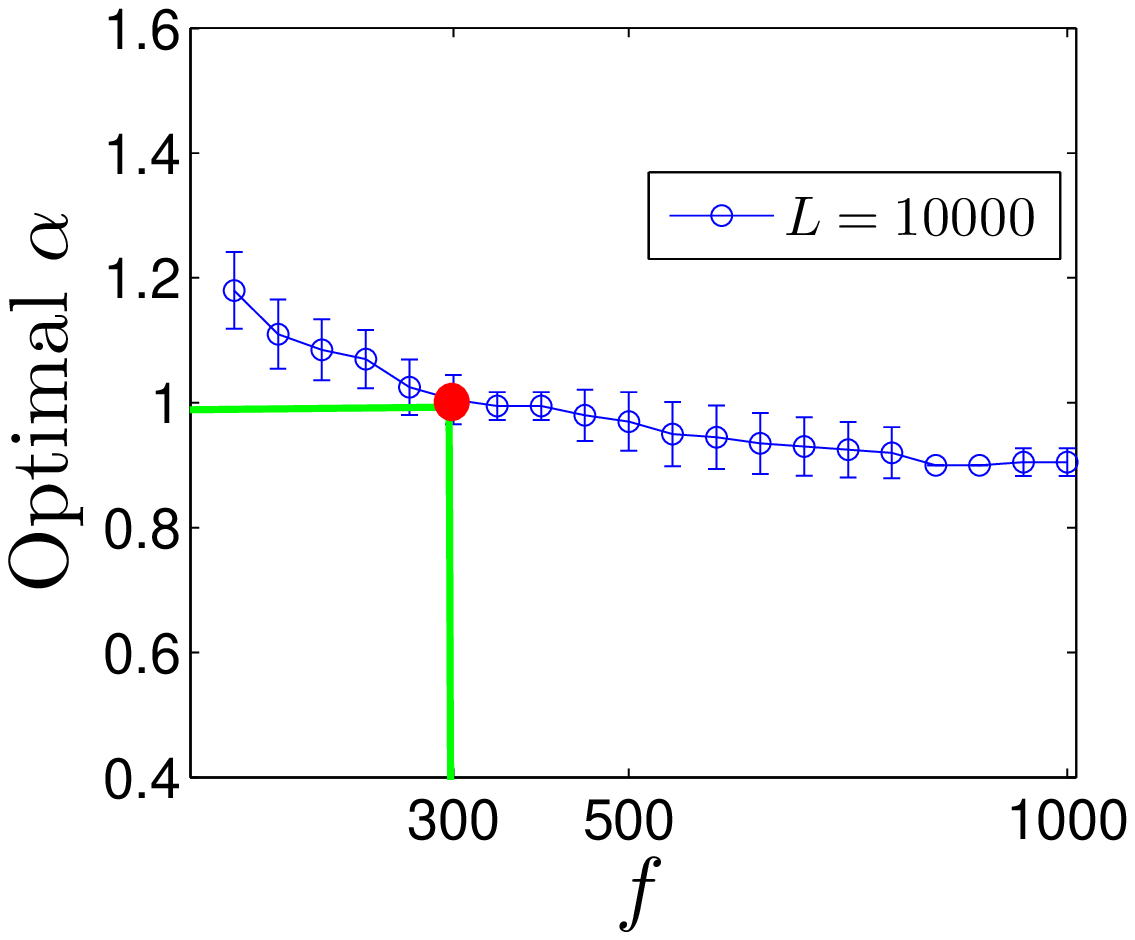}\includegraphics[width=4.4cm,height=3.8cm]{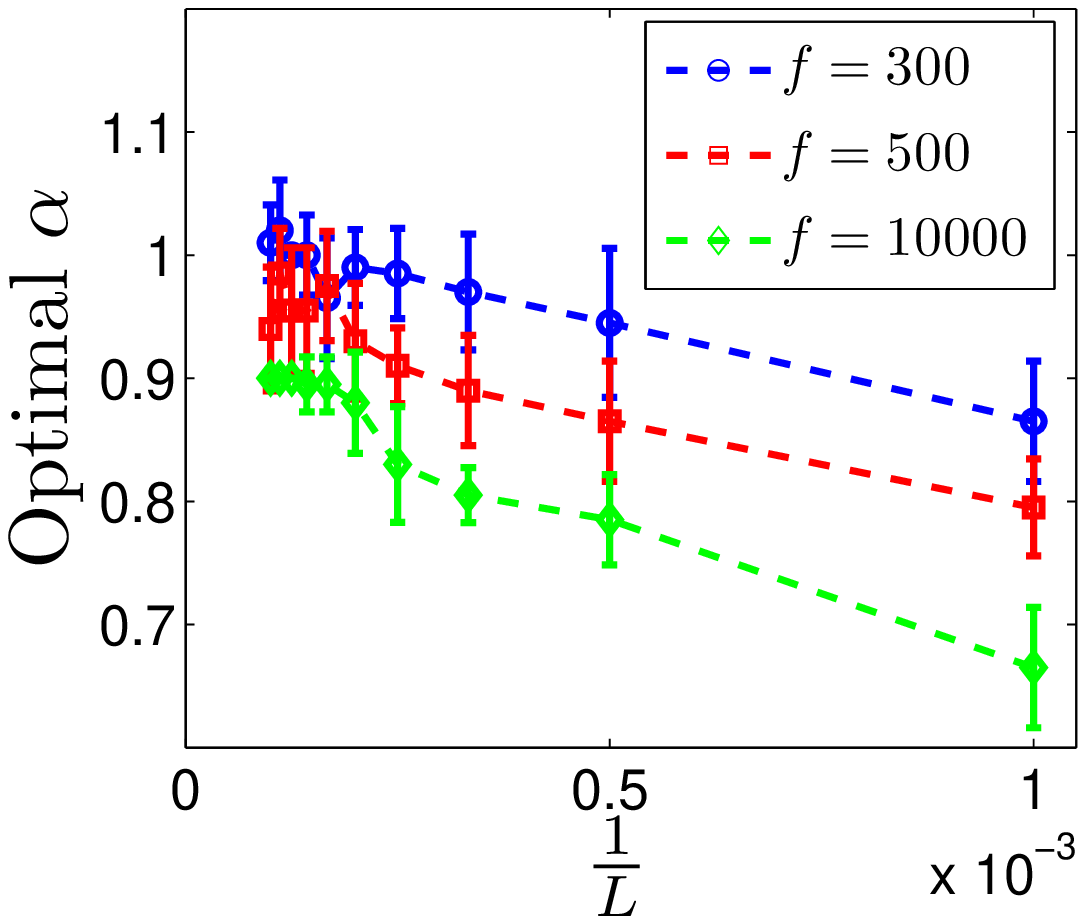}
\vskip -23.5em \hskip -27.0em  \raisebox{1em}{\bf{a}} \vskip -1.9em
\hskip 1.8em {\bf{b}} \vskip 10.5em \hskip -26.5em{\bf{c}} \vskip
-1.0em \hskip 1.8em {\bf{d}}\vskip 9.0em \hskip -20.0em \caption{The
relationship between $\varepsilon, w, f$, $\alpha$ and $L$ in the
optimization model. \textbf{a}. The contour map shows the
relationships between $w$, $\alpha$ and $\varepsilon$, for
$L=10000$. The colors indicate the value of $\varepsilon$. In
\textbf{b}, the dependence of the information entropy $\varepsilon$
on $\alpha$ for $f=300, 500, 1000$ is shown. \textbf{c}. The
dependence of the optimal $\alpha$ on the average number of friends
$f$. The error bars denotes the standard deviations. \textbf{d}. The
relationships between optimal $\alpha$ and the edge length $L$ of
the lattice. From it we can see that for large $L$ the optimal
$\alpha$ approaches $1$. The error bars denotes the standard
deviations.} \label{optimal}
\end{figure}

For a large enough 2-dimensional lattice, the number of nodes that
have distance $r$ from a given node is proportional to $r$. So if
$w\rightarrow +\infty$, that means if we consider the maximal
diversity of friendships without any constraints of energy, we
expect $P(r)\propto r$ to be the optimal entropy information since
each node has the same probability in the information sequence. In
practice, individuals naturally have a limited energy $w$. Our
numerical results shown in Fig. \ref{optimal}.\textbf{a} indicate
that when $\alpha \approx1$, the information entropy $\varepsilon$
is near its maximum value for a very broad range of $w$. For the
range $w\in(5\times10^4, 10^6)$ and $f\in(50, 1000)$, we find the
optimal $\alpha$ to be $\alpha=1\pm0.05$.

When the size of the lattice is $L$ and $P(r)\propto r^{-1}$, the
mean distance between friends is $\frac{L}{\log{L}}$. Therefore, we
can find the average number of friends $f$ to be
\begin{equation}f=\frac{w\log L}{L}\label{wf}\end{equation}
which gives one to one correspondence between $f$ and $w$ at the
optimal state. When $L=10000$ and $w\in(5\times10^4, 10^6)$ the
average number of friends is $f\in(50, 1000)$ which indeed
corresponds to reality \cite{number_friends}. In particular, when
considering the average number of friends we contact in one year,
$f=300$ \cite{number_friends}, the optimal value of $\alpha$ is
$\alpha=-0.99\pm0.03$ (as shown in Fig. \ref{optimal}).

\begin{figure}
\center
\includegraphics[width=4.4cm,height=3.8cm]{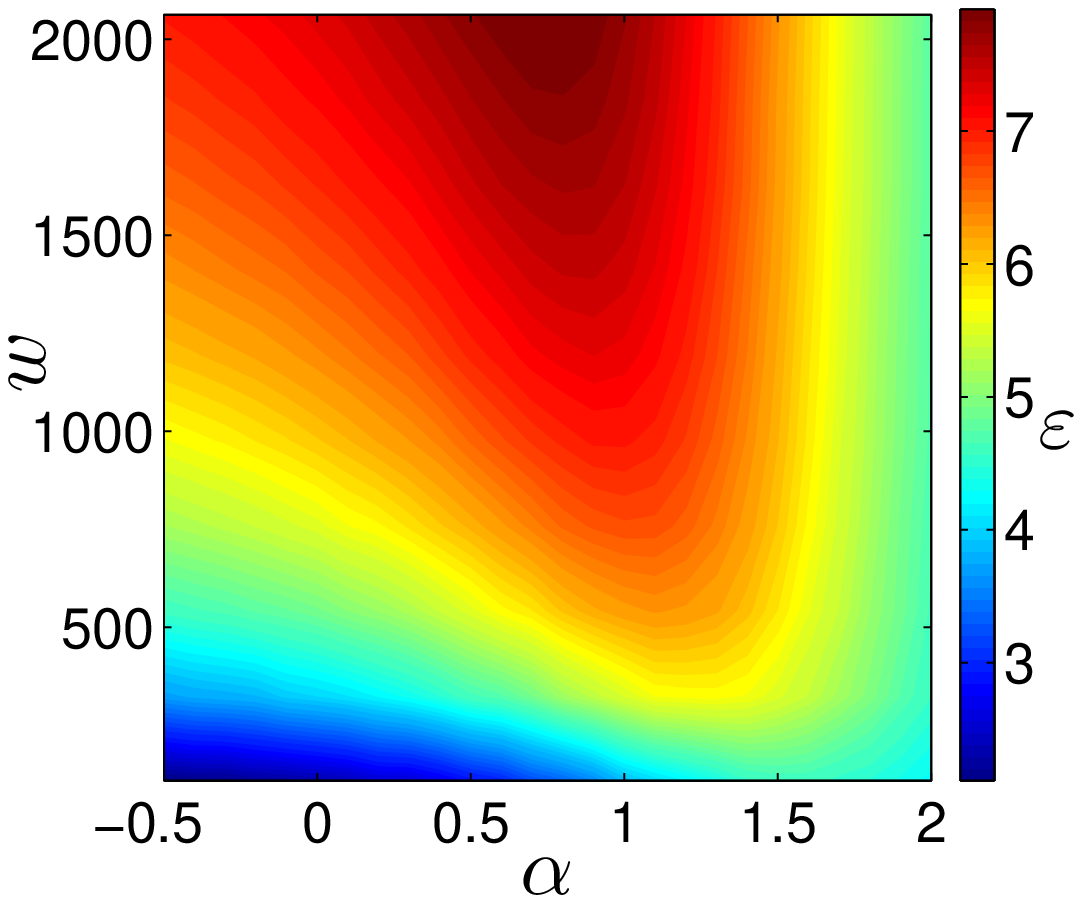}\includegraphics[width=4.4cm,height=3.8cm]{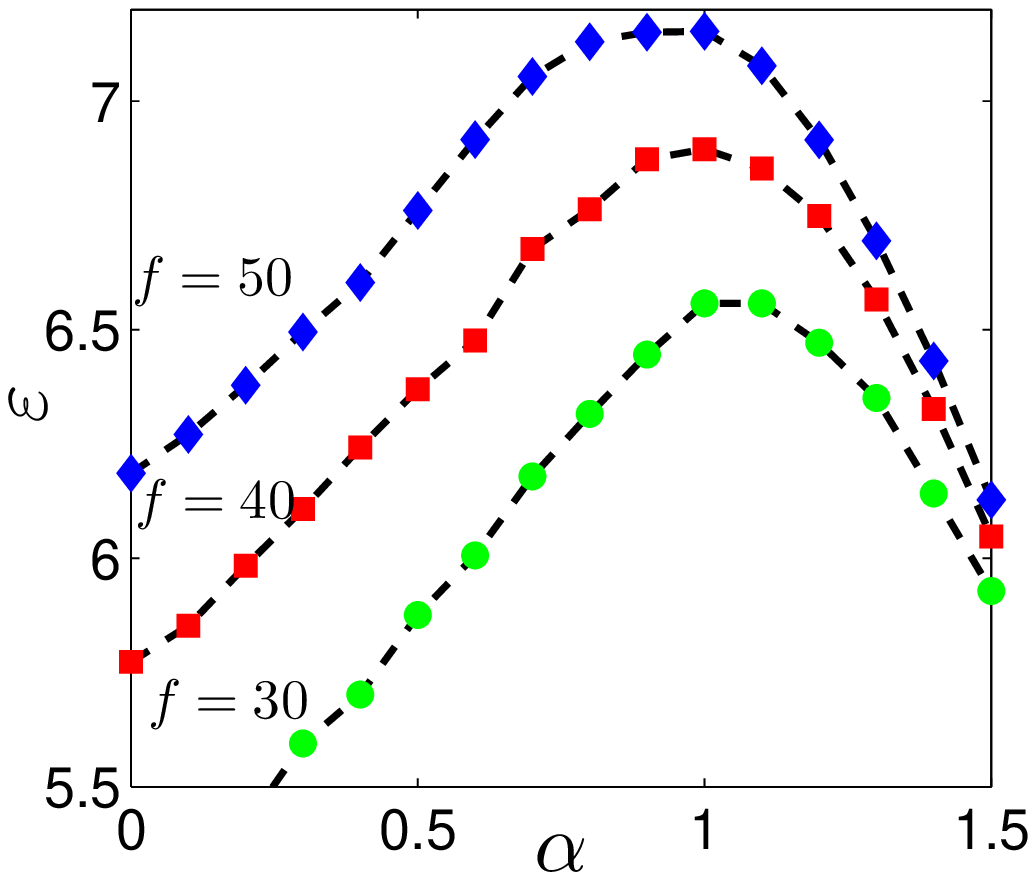}
\includegraphics[width=4.4cm,height=3.8cm]{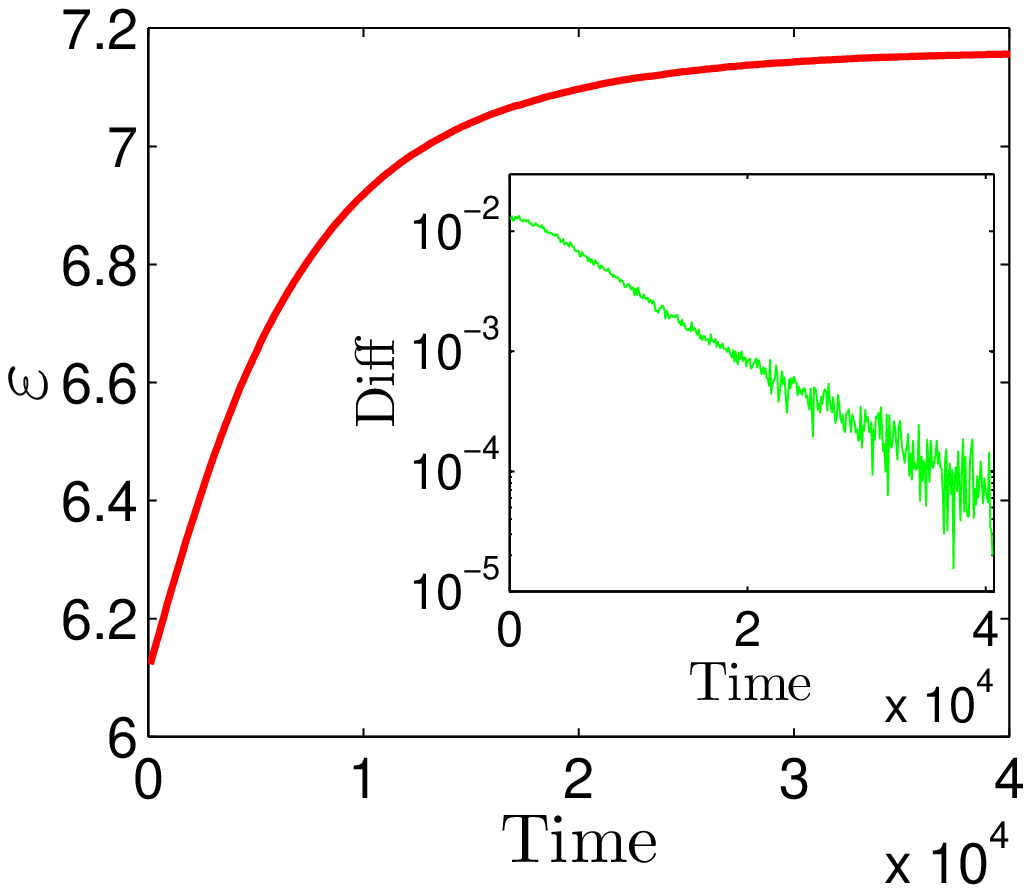}\includegraphics[width=4.4cm,height=3.8cm]{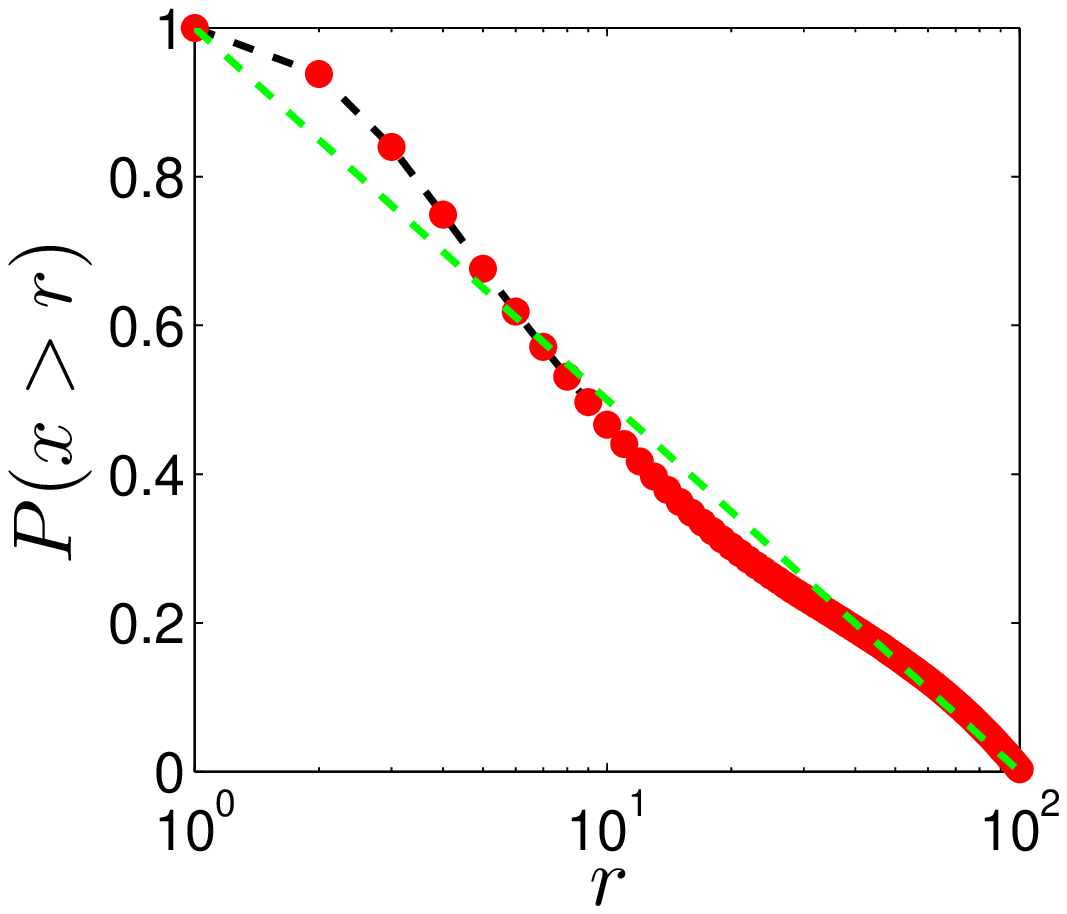}
\vskip -23.5em \hskip -27.0em  \raisebox{1em}{\bf{a}} \vskip -1.9em
\hskip 1.8em {\bf{b}} \vskip 10.5em \hskip -26.5em{\bf{c}} \vskip
-1.0em \hskip 1.8em {\bf{d}}\vskip 9.0em \hskip -20.0em \caption{The
results of evolutionary model when $L=100$ and $f=50$. \textbf{a}.
The simulation results of OM on a toroidal lattice with the preset
power law distribution $P(r)\propto r^{-\alpha}$. \textbf{b}. the
dependence of the information entropy $\varepsilon$ or $\alpha$ for
$f$ around 40 in the OM. We can see that when $f=50$, the optimal
exponent is 0.95 and it is very close to $-1$. \textbf{c}. The
changes of entropy in the EM with the evolution time. The entropy is
fixed and the system archives a steady state. The fixed entropy is
7.15 which is very close to the entropy 7.18 in the network of
$L=100$ which we preset the distribution is $P(r)\propto r^{-1}$.
The inset denotes the difference of the time-entropy curve which
implies that the difference decays exponentially. From it we can see
that for a sufficient long time evolution, the entropy converges to
a fixed value and the system achieves a steady state. \textbf{d}.
The cumulative distribution of the distance in EM is shown in
log-linear plot in the steady state. We can see that this
distribution is very close to $P(r)\propto r^{-1}$ (dashed
line).}\label{evolution}
\end{figure}

Our results suggest that $P(r)\propto r^{-1}$ is the optimal
distribution for collecting information between all power law
distributions. Is $P(r)\propto r^{-1}$ the optimal distribution when
considering all kinds of distributions? We will demonstrate, based
on the following evolutionary model (EM), that among all kinds of
distributions, $P(r)\propto r^{-1}$ is still the optimal one. In the
EM, we also construct a network on a lattice of size $L\times L$. A
node $u_i$ is one of the neighbors of node $u$ when there is a
direct link from $u$ to $u_i$. Each node $u$ has friends at
distances $r(u,u_i)$ subject to $\sum_{u_i\in U}r(u,u_i)\leq w$,
where $U$ is the set of all neighbors of node $u$. In the initial
stage of the EM, $P(r)$ is set to be a uniform distribution. Then we
employ the extremal optimization method
\cite{Extremal_optimization}, to maximize the entropy through the
evolution of network architecture. At each step, a node is chosen
randomly. For a chosen node $u$, we make two operations, deleting
and adding neighbors according to the marginal improvement of
entropy. Suppose $u$ has $k$ neighbors. For the deleting execution,
we first calculate the marginal entropies of each neighbor of node
$u$, $\{\frac{\vartriangle E_{u_1}}{r(u,u_1)},\frac{\vartriangle
E_{u_2}}{r(u,u_2)}, \cdots, \frac{\vartriangle
E_{u_{k}}}{r(u,u_{k})}\}$, where $\vartriangle E_{u_i}$ means the
change in the entropy of node $u$ when we delete node $u_i$ from the
neighborhood of node $u$ with other parameters being unchanged. Then
we randomly select a comparatively small $|\frac{\vartriangle
E_{u_i}}{r(u,u_i)}|$ with probability $Pr(u_i)$ proportional to
$(rank |\frac{\vartriangle E_{u_i}}{r(u,u_i)}|)^{-1-log(k)}$
\cite{Extremal_optimization} and delete $u_{i}$ from $u$'s
neighborhood. For the adding link execution, suppose $v_1, v_2,
\cdots, v_h$ are all the candidates which are currently next nearest
neighbors of node $u$. We first calculate the marginal entropies of
each of the candidate, $\{\frac{\vartriangle
E_{v_1}}{r(u,v_1)},\frac{\vartriangle E_{v_2}}{r(u,v_2)}, \cdots,
\frac{\vartriangle E_{v_h}}{r(u,v_h)}\}$, then we also employ the
extremal optimization method to choose a node whose marginal entropy
is comparatively large among all candidates' marginal entropies as a
friend of node $u$. We repeat the adding execution until all the
candidates are chosen or the energy limit (Eq. (2)) is satisfied.

In the evolutionary model, we have to record all friends of each
node and therefore a system of size $L\times L$ with $L=10000$ is
too large to simulate. So we simulate the evolutionary model on a
toroidal lattice of size $100\times 100$. We assume that the energy
scales linearly with distance as suggested by Eq. (2). Thus, when
reducing $L$ from 10,000 to 100 (factor of 100) we expect the
corresponding energy to be reduced from order of $10^5$ to order of
$10^3$. We therefore study the EM model of $L=100$ with $w=1086$
$(f=50)$.

In order to find the optimal distribution of the distances, we first
employ the optimization model described by Eqs. (2)-(4) to analyze
the above case with the system size $100\times 100$ and
$w\simeq10^3$. We find that the maximum entropy is 7.18 and the
corresponding $\alpha$ is $\alpha=0.95\pm 0.05$ (see Fig.
\ref{evolution}a, b). Next we simulate the evolutionary model of
size of $100\times 100$ and $w\simeq10^3$. After long term evolution
from the initial uniform distribution (each node modify the
neighborhood more than 40000 times), the system achieves its
stationary state (Fig. \ref{evolution}e). The maximum entropy is
7.15 and the corresponding PDF of the distance between the friends
scales as $P(r)\propto r^{-1}$ (Fig. \ref{evolution}d and
supplementary IIIB), which are very close to the results obtained by
OM. So we conclude that $P(r)\propto r^{-1}$ is the optimal PDF of
distances of friendships for collecting maximal information. It
implies that, the spatial structure of the real social networks is
the most optimal structure which leads to the maximum diversity of
the friends' location and can help individuals to collect
information efficiently. We note that, it can be proved
analytically, under the assumption that the energy scales linearly
with system size, i.e. $w=cL$, for $L\rightarrow +\infty$, that
$P(r)\propto r^{-1}$ will be the optimal distribution for maximizing
entropy among all power law distributions (see supplementary IIC for
detailed analysis ).
\section{Conclusion}
From the empirical results, we conclude that the probability
distribution of having a friend at distance $r$ scales as
$P(r)\propto r^{-1}$ which is a universal spatial property for
social networks. It is shown here that the origin of this spatial
scaling law may result from the maximization of entropy which can
benefit individuals for optimally collecting information.
\\
\\

\textbf{Acknowledgement.} We thank Dr. R. Lambiotte for providing us
the mobile network data at the beginning of this research. Yanqing
Hu wishes to thank Prof. Fukang Fang, Prof. Gang Hu, Dr. Jiang
Zhang, Dr. Erbo Zhaofor and Dr. Yiming Ding for very useful
discussions. The work is partially supported by NSFC under Grant No.
70974084, No. 60534080 and No. 70771011. Daqing Li and Shlomo Havlin
thank the ONR, EU project Epiwork and the Israel Science Foundation
for financial support.

\textbf{Supplementary Information}

\section{Explanation for Spatial Scaling of LiveJournal }
In the empirical study of the LiveJournal data set \cite{Use
Kleinberg search1}, for each distance $r$, $Q(r)$ is the fraction of
friendships among all pairs $u$, $v$ of LiveJournal users with $r(u,
v)=r$. $Q(r)=\frac{F(r)}{S(r)}\propto r^{-1}$. Here, $F(r)$ denotes
the total number of friendships with distance $r$ and $S(r)$ is the
total number of pairs of nodes that have distance $r$. The
LiveJournal social network has a fractal dimension of about $0.8$
(they define the fractal dimension of a network as the exponent $d$
of the best-fit function $rank_u(v)=c\cdot r(u, v)^d$, where
$rank_u(v)$ is the number of people who live closer to $u$ than $v$
and $c$ is a constant). We know that for any $d$-dimensional
lattice, the number of nodes that have the same distance $r$ to a
given node is proportional to $r^{d-1}$. In fractal networks, $d$
should be the fractal dimension. Thus the probability density
function $P(r)$ of the geographic distance $r$ between friends is
about $P(r)\propto r^{d-1}\cdot Q(r)=r^{0.8-1}\cdot r^{-1}=r^{-1.2}$
, which is close to $r^{-1}$.

\section{About the Optimization Model (OM)}

\subsection{Why We Only Consider Friends and Next Nearest Friends?}

\begin{figure}
\center
\includegraphics[width=7cm]{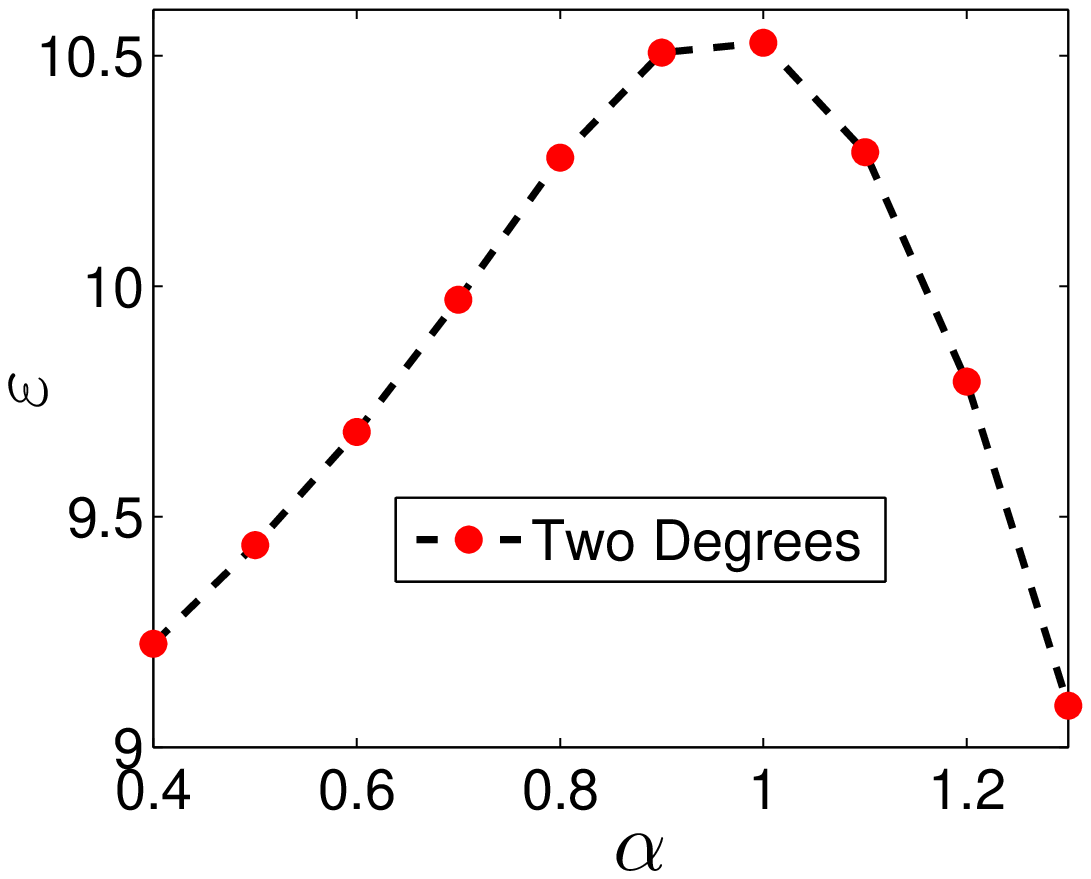}\includegraphics[width=7cm]{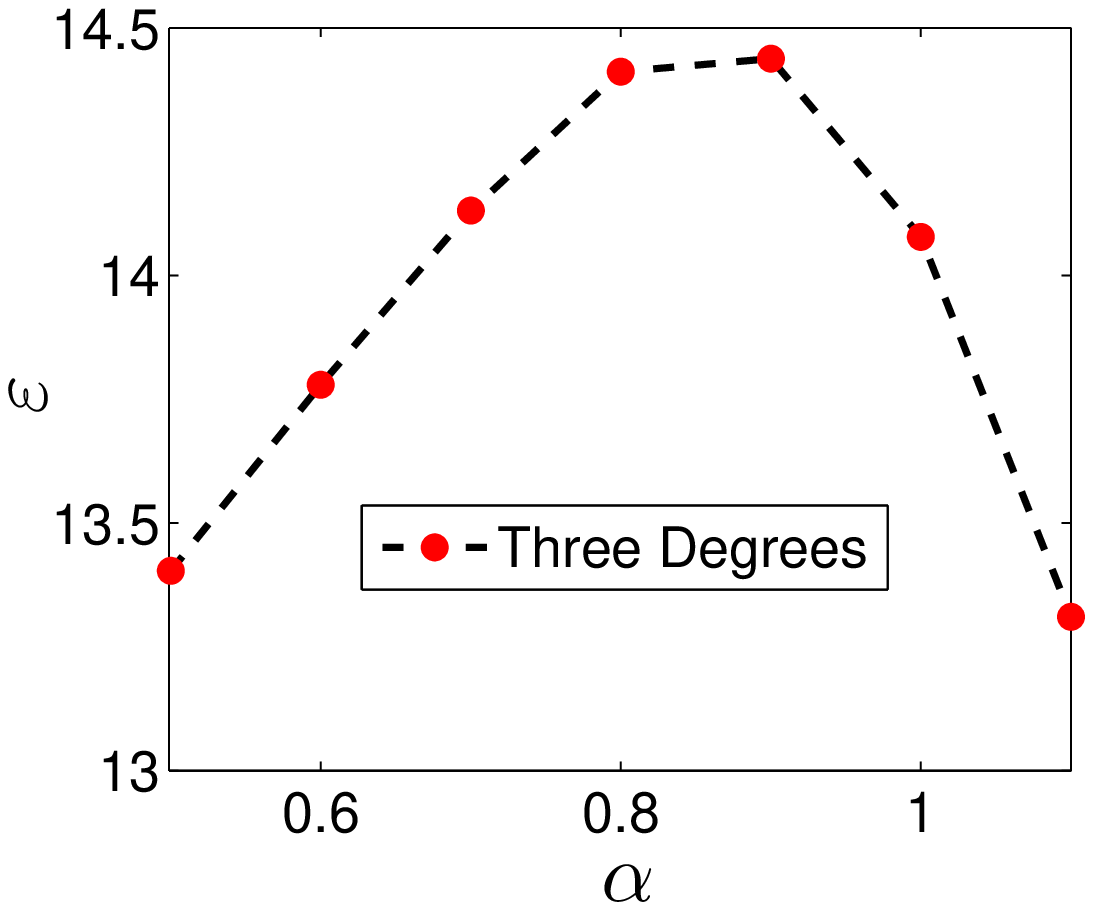}
\caption{The relationship between entropy and the power law exponent
in different degrees of influence. The lattice size is
$3000\times3000, f=300$. We can see that the phenomena are similar
in which -1 is close to the optimal exponents.}\label{friends-rank}
\end{figure}

We assume that the information obtained from the social network is
actually related with the influence of friendships. Indeed,  in our
social life, our friends always talk something about their friends.
Thus, we assume that friends and next nearest friends are most
important and is enough to consider them in our model. However,
Christakis and Fowler have found recently that the influence is
mainly within three degrees of separation and call this finding the
``Three Degrees of Influence Rule" \cite{Fowler}. It is
computationally difficult to take into account more than two degrees
of separation of friends to study a system of $10^4\times10^4$. We
have therefor performed the numerical experiments of the OM in
$3000\times3000$ size lattice with $w\simeq10^4$ ($f=300$) and found
that the simulated results were similar when we took into account
friends and next nearest friends, and three degrees of separation
(as shown in Fig. \ref{friends-rank}).

\subsection{Algorithm of OM}

When the lattice size is $10000\times10000$, it is hard to record
all nodes' links information. Thus, we first represent each node an
index running from 1 to $10^8$. This way is easy to obtain a
function $r(u,v)$ to calculate the lattice distance between any pair
of nodes $u$ and $v$, where $u, v$ are now the running index.

In the OM model all nodes are equivalent. Without losing generality,
we can set any node as $u=1$. To construct the spatial network on
the lattice, each time we first randomly generate a distance $r$
according to the distribution $P(r)\propto r^{-\alpha}$,
$r\in\{1,2,\cdots L\}$. Then from the set of nodes which have
distance $r$ from node $1$, a node is chosen randomly as a friend of
node $1$ and a directed link is constructed. Repeating the execution
until the energy achieves the limit constraint. After the executions
we can get all the friends of node $1$. Employing the same approach,
we can also get all the next nearest friends of node $1$.

\subsection{Analysis on OM} In this section we will prove that if
energy hold
\begin{equation}w=cL,\end{equation} where $c$ is constants,
for $L\rightarrow+\infty$, $P(r)\propto r^{-1}$ is the optimal
distribution for all $P(r)\propto r^{-\alpha}$ distributions.

\subsubsection{Symbol and Expression Descriptions}

$P(r)\propto r^{-\alpha}$, the distribution of distance between
friendships.

$R_{\alpha}$ , the expectation of the distance which holds
$P(r)\propto r^{-\alpha}$.

$f_\alpha=\frac{w}{R_{\alpha}}$ is the expectation of number of
friends.

When $w=\frac{fL}{\log(L)}$, $L$ is the edge length of the lattice,
$f$ denotes the number of friends when $\alpha=1$.

$q_{i,j}^{\alpha}$ denotes the probability of the connection between
node $i$ and $j$ for a given $\alpha$.

$F^{\alpha}=\{\theta_1^{\alpha},
\theta_2^{\alpha},\cdots,\theta^{\alpha}_{f_{\alpha}}\}$, denotes
the set of friends of node 1, where $f_{\alpha}=\frac{w}{R_\alpha}$.

$q_{F^{\alpha},i}=\frac{1}{f_\alpha}\sum_{v=1}^{f_{\alpha}}q^{\alpha}_{\theta^{\alpha}_v,i}$,
denotes the probability that node $i$ is one of friends of
$F^\alpha$.

$\sum_{x=1}^{f_\alpha^2}\frac{x}{f_\alpha^2}\log\frac{x}{f_\alpha^2}C_{f_\alpha^2}^xq_{{F^{\alpha}},i}^x(1-q_{{F^{\alpha}},i})^{{f_{\alpha}^2}-x}$
denotes the expectation of entropy of node $i$ when the chosen
probability of node $i$ is $q_{F^{\alpha},i}$ and the time of
choosing is ${f_\alpha^2}$.

$\varepsilon_{\alpha}=\sum_{i=1}^n\sum_{x=1}^{f_\alpha^2}\frac{x}{f_\alpha^2}\log\frac{x}{f_\alpha^2}C_{f_\alpha^2}^xq_{{F^{\alpha}},i}^x(1-q_{{F^{\alpha}},i})^{{f_{\alpha}^2}-x}$,
denotes the expectation of entropy for a given ${F^{\alpha}}$.

$E(\varepsilon_{\alpha})$, denotes the expectation
$\varepsilon_{\alpha}$

\subsubsection{Case 1: $\alpha<1$}

\begin{equation}R_{\alpha}=\frac{\int_{1}^{L}x^{1-\alpha}dx+O(1)}{\int_{1}^{L}x^{-\alpha}dx+O(1)}=\frac{\frac{1}{2-\alpha}(L^{2-\alpha}-1)+O(1)}{\frac{1}{1-\alpha}(L^{1-\alpha}-1)+O(1)}\approx\frac{1-\alpha}{2-\alpha}L.\end{equation}

Therefore, for a given $w=cL$, where $c$ is a constant, we have

\begin{equation}\lim_{L\rightarrow\infty}f_{\alpha}=\lim_{L\rightarrow\infty}\frac{w}{R_{\alpha}}=\frac{c(2-\alpha)}{1-\alpha}.\label{cas1f}\end{equation}

Because, \begin{equation}\lim_{L\rightarrow\infty}\max
q^{\alpha}_{i,j}\leq
\lim_{L\rightarrow\infty}\frac{1}{\int_{1}^{L}x^{-\alpha}dx+O(1)}=\lim_{L\rightarrow\infty}\frac{1}{\frac{1}{1-\alpha}(L^{1-\alpha}-1)+O(1)}=0\end{equation}
and
\begin{equation}q_{{F^{\alpha}},i}\leq \max q^{\alpha}_{i,j}.\end{equation}
Thus, for any ${F^{\alpha}}$,
\begin{equation}\lim_{L\rightarrow\infty}q_{{F^{\alpha}},i}=0.\end{equation}
It implies that
\begin{equation}\lim_{L\rightarrow\infty}\varepsilon_{\alpha}=\log(\frac{c(2-a)}{1-a}+[\frac{c(2-a)}{1-a}]^2).\end{equation}
Thus
\begin{equation}\lim_{L\rightarrow\infty}E(\varepsilon_{\alpha})=\log(\frac{c(2-a)}{1-a}+[\frac{c(2-a)}{1-a}]^2),\end{equation}
which is a monotonic increasing function with $\alpha<1$.

\subsubsection{Case 2: $\alpha>1$}

\begin{figure}
\center
\includegraphics[width=8cm]{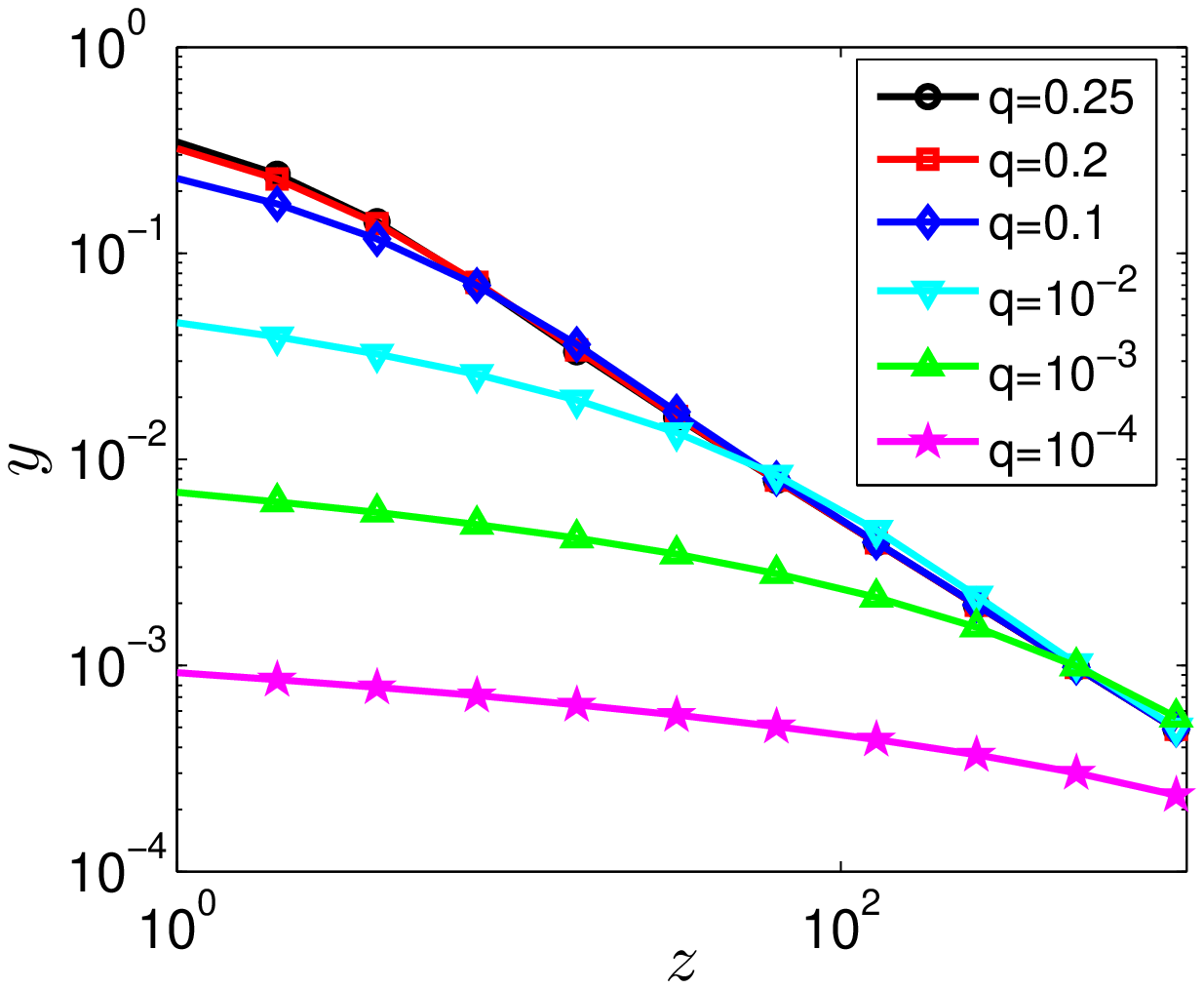}\includegraphics[width=8cm]{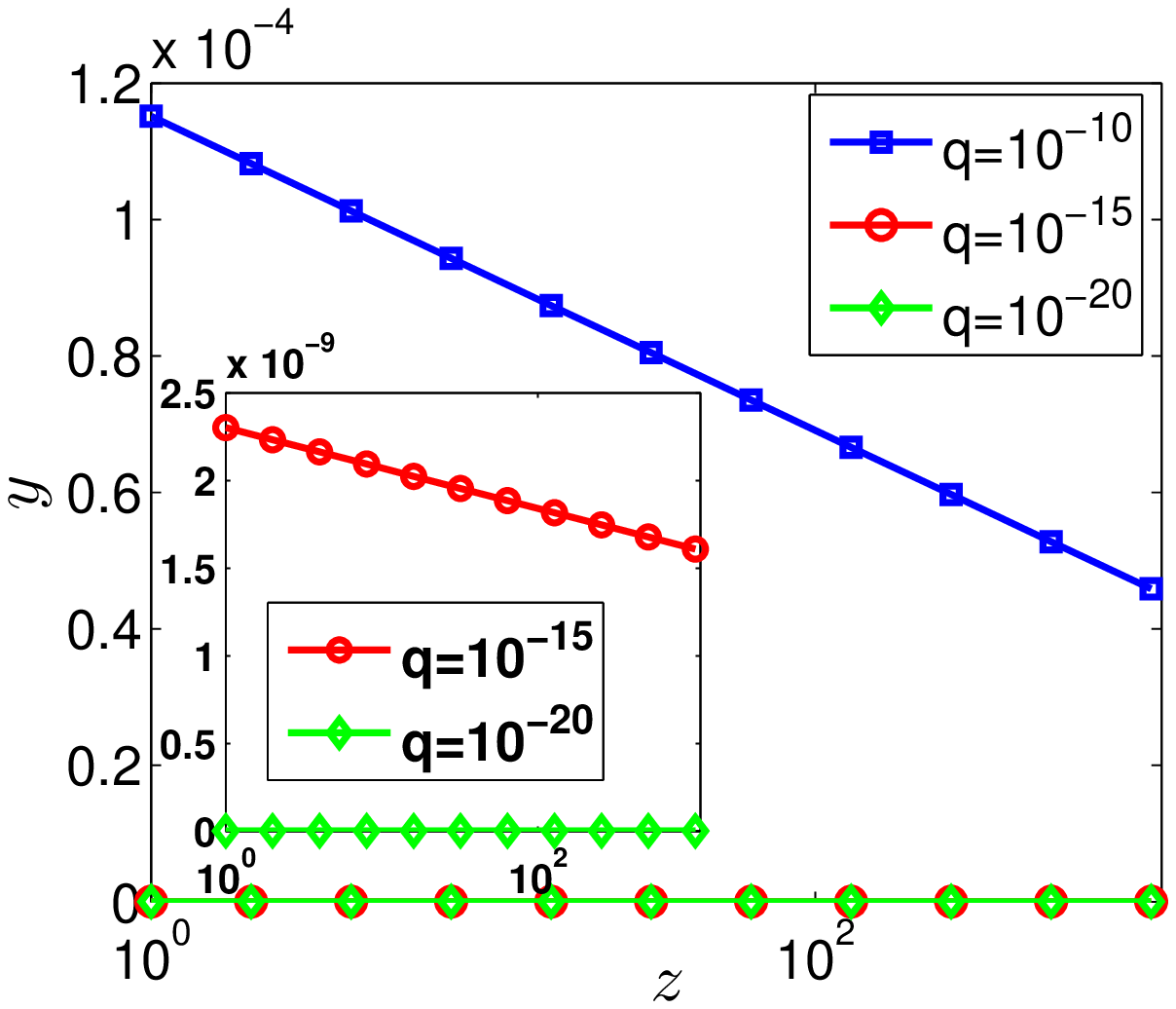}
\caption{Plot of $y=-q\log q-g(z)$. From the plot we can see that
Lemma is true. More over when $z$ is small $g(z)>-q\log q$ is also
correct.}\label{fact2}
\end{figure}

\textbf{Lemma}:  if $q\in(0,\frac{1}{3})$, for any large enough $z$
we have
\begin{equation}-q\log
q>-\sum_{x=1}^{z}\frac{x}{z}\log\frac{x}{z}C_{z}^xq^x(1-q)^{{z}-x},\end{equation}
where
$-\sum_{x=1}^{z}\frac{x}{z}\log\frac{x}{z}C_{z}^xq^x(1-q)^{{z}-x}$
denotes the expectation of entropy of a node with the probability
$q$ to be chose and the total choosing time is $z$ (as shown in Fig.
\ref{fact2}).

\textbf{Proof}:

According to Law of Large Numbers,
$\lim_{z\rightarrow\infty}-\sum_{x=1}^{z}\frac{x}{z}\log\frac{x}{z}C_{z}^xq^x(1-q)^{{z}-x}=-q\log
q$.

Thus, we just need to prove
\begin{equation}g(z)=-\sum_{x=1}^{z}\frac{x}{z}\log\frac{x}{z}C_{z}^xq^x(1-q)^{{z}-x}\end{equation}
is a monotonic increasing function.

For large enough $z$,  normal distribution is a well approximation
to binomial distribution then we have
\begin{equation}g(z)=\int_1^z\frac{1}{\sigma\sqrt{2\pi}}e^{\frac{-(x-\mu)^2}{2\sigma^2}}\frac{x}{z}\log\frac{x}{z}dx,\end{equation}
where $\sigma^2=zq(1-q), \mu=zq$.

\begin{equation}g_z^{'}(z)=\frac{\sqrt{2}}{4z^3q(1-q)\sqrt{\pi zq(1-q)}}\int_1^z[\log\frac{x}{z}(q^2z^2-3q^2z+3qz-x^2)-2zq^2+2zq]e^{\frac{(x-zq)^2}{2zq(q-1)}}xdx\end{equation}

Obviously, \begin{equation}\frac{\sqrt{2}}{4z^3q(1-q)\sqrt{\pi
zq(1-q)}}>0\end{equation} and
\begin{equation}e^{\frac{(x-zq)^2}{2zq(q-1)}}x>0\end{equation}.

More over
\begin{equation}\int_1^z[\log\frac{x}{z}(q^2z^2-3q^2z+3qz-x^2)-2zq^2+2zq]dx=(\frac{1}{9}-q^2)z^3+\Theta(z^2\log z)>0\end{equation}
when $q<\frac{1}{3}$, where, $\Theta(z^2\log z)$ denotes the same
order of $z^2\log z$.

Thus, $g_z^{'}(z)>0$ which implies that $g(z)$ is a monotonic
increasing function and \begin{equation}-q\log
q>-\sum_{x=1}^{z}\frac{x}{z}\log\frac{x}{z}C_{z}^xq^x(1-q)^{{z}-x}.\end{equation}

For case 2, according to Lemma and Levy stable distribution property
(the distance between the next nearest neighbor and the origin is
also obey $P(r)\varpropto r^{-\alpha}$ when $\alpha>1$). So for
large enough friends number we have:
\begin{equation}
E(\varepsilon_{\alpha})<\sum_{r=1}^{L}4r\frac{r^{-\alpha}}{4rZ(\alpha)}\log\frac{r^{-\alpha}}{4rZ(\alpha)}\end{equation}
\begin{equation}=\frac{1}{Z(\alpha)}\sum_{r=1}^{L}r^{-\alpha}\{(-\alpha-1)\log
r-\log[4Z(\alpha)]\}\label{alq2}.\end{equation} More over we can
get:
\begin{equation}\lim_{L\rightarrow+\infty}E(\varepsilon_{\alpha})=\frac{(a-1)(2\log2+\log Z(a))+a+1}{2(a-1)^2}\label{case3}\end{equation}
where $Z(\alpha)$ denotes $\sum_{r=1}^{L}r^{-\alpha}$. Obviously,
$\frac{(a-1)(2\log2+\log Z(a))+a+1}{2(a-1)^2}$ is a monotonic
increasing function. Thus, for any fixed $c$, -1 is the optimal
exponent.

\section{About the Evolutionary Model (EM)}

\subsection{Why we chose new friend only from the next nearest neighbors?}
There are 2 reasons. The first is that, according to our real social
experience, we always make some new friends who are the friends of
our friends. The second is that EM is a global optimal algorithm.
Thus if we choose any node as our new friend, the result will be the
same theoretically.

\subsection{How to Measure the Power Law Exponent in EM?}
To accurately measure the exponent value of power law distribution
is not a easy work. Especially, when the exponent is very close to
$-1$. We use the least square method to evaluate the exponent value.
We are afraid the least square method is not a good way, so we plot
the accumulated curve. Fortunately, it can be proved that when
$P(r)\propto r^{-1}$, the accumulated function in log-linear plot
will be a straight line. We can see that the distribution is about
$P(r)\propto r^{-1}$.

\end{document}